# Analysis of Electric Vehicle Charging Station Usage and Profitability in Germany based on Empirical Data

Christopher Hecht[*, 1, 2, 3]
https://orcid.org/0000-0002-2763-659X

Jan Figgener[1, 2, 3]
https://orcid.org/0000-0003-2216-9432

Dirk Uwe Sauer[1, 2, 3, 4]
https://orcid.org/0000-0002-5622-3591

[1]*Institute for Power Electronics and Electrical Drives, RWTH Aachen University, Germany*
[2]*Institute for Power Generation and Storage Systems, RWTH Aachen University, Germany*
[3]*Juelich Aachen Research Alliance, JARA-Energy, Germany*
[4]*Helmholtz Institute Muenster (HI MS), IEK-12, Forschungszentrum Jülich, German*

*Abstract*—Electric vehicles are booming and with them the required public charging stations. Knowing how charging stations are used is crucial for operators of the charging stations themselves, navigation systems, electricity grids, and many more. Given that there are now 2.5 as many vehicles per charging station compared to 2017, the system needs to allocate charging points intelligently and efficiently. This paper presents representative data on energy consumption, arrival times, occupation, and profitability of charging stations in Germany by combining usage data of 27,800 installations. Charging happens mainly during the day and on weekdays for AC charging stations while DC fast-charging stations are more popular on the weekend. Fast-chargers service approximately 3 times as many vehicles per connection point while also being substantially more profitable due to higher achieved margins. For AC chargers, up to 20 kWh of energy are charged in an average charge event while DC fast-chargers supply approximately 40 kWh.

*Keywords—Charging stations, electric vehicles, usage, profitability, Germany*

## I. Introduction

Global warming is a key challenge of the 21st century. To tackle this issue, greenhouse gas emissions must be reduced across all sectors of human activity. While progress has been made in many sectors, the traffic sector is particularly hard to decarbonize and emissions have risen by 74% between 1990 and 2016 worldwide [1]. One approach to do so is to electrify transport. For road-bound electric vehicles (EVs), this requires public recharging opportunities (see Figure 2 for a component overview). Such public charging stations (PCS) are expensive to install. To make such investments worthwhile, good knowledge of PCS usage is required.

Other parties with an interest in PCS use patterns are navigation system providers, grid operators, energy system modellers, and many more. Navigation system providers require this information to ensure that drivers find a free recharging opportunity either on the way or upon arriving at their destination. Grid operators need to dimension electricity grids supplying PCSs to a realistic demand. Energy system modellers need to understand how much energy is required when and where for power generation and storage to meet demand.

These questions gain an increasing urgency worldwide. Taking Germany as a use case, the number of battery electric vehicles (BEVs) that have to share a single PCS is steadily increasing (see Figure 1) reaching approximately 25 BEVs per PCS by the end of 2021. To avoid excessive queuing at PCS, all actors mentioned need to increase the intelligence of their system.

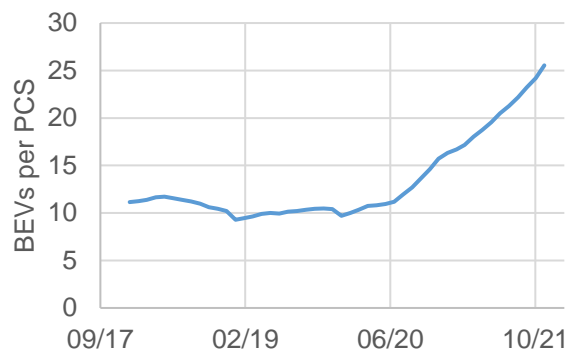

*Figure 1: BEVs per PCS in Germany [2, 3]*

In this paper, we provide a detailed analysis of how, where, and when PCSs are used. Using a dataset comprised of 27,800 PCSs in Germany observed between 2019 and 2021 the study is unmatched in terms of amount and representativeness [4–12]. Specifically, we address the following research questions:

1. How much energy is recharged per charge event?
2. When people arrive at PCSs?
3. How high is the occupation of PCSs?
4. How long do vehicles stay at PCSs?
5. How profitable are PCSs?

To give readers a holistic answer to all of the questions above, results are shown along several dimensions. Individual Electric Vehicle Supply Equipment (EVSE, one EVSE can supply one car, but may have multiple connectors and multiple EVSEs form one PCS) units are categorised by:

1. The rated power of connectors (connectors are cables or sockets to connect vehicle and PCS)
2. The type of connectors
3. The type of land that EVSEs are placed on

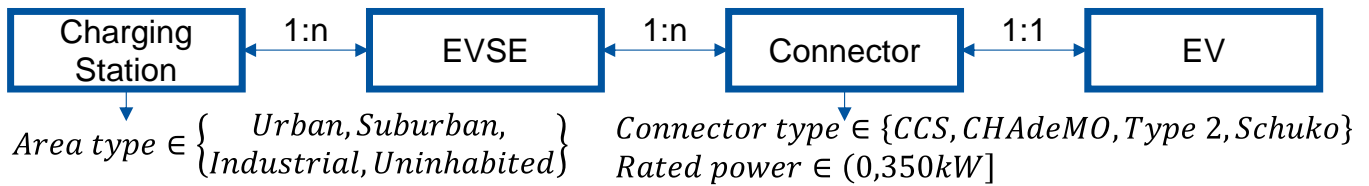

*Figure 2: Relationship between the components and properties of a (public) charging station in the context of this work.*

## A. Literature review

In the early days of PCS deployment, little empirical data was available [13–16]. Scholars therefore had to use traffic counter data assuming that a share was electric vehicles [17, 18], models based on socio-economic data and expert opinions [19, 20]. The challenge with the two options is that although large datasets are available, the link between the data and the actual PCS usage is hard to prove. When using electric vehicle field data, a typical approach is to assume that vehicles charge where they are parked. This allows deriving a model of where PCSs would be most occupied. The issue with this approach is that monitored vehicle fleets are typically quite small and representativeness cannot necessarily be assumed.

With the increasing market share of electric vehicles in many countries, researchers took a new approach: large datasets from PCS usage started to become available. Works were performed about Amsterdam and its surrounds [7, 8], the Netherlands as a whole [9–11], Nebraska, USA [4–6], and Jiangsu Province, China [12]. The level of detail differs between each study, but generally, at least the occupation status per PCS and in some cases the energy per charge event were known. Note that the authors of the studies are overlapping indicating that fewer datasets are available than papers published. An overview over which study contains what kind of information can be found in [21] and [22]. A key remaining issue is representativeness, particularly if the area under observation contains only a small population such as Amsterdam or Nebraska.

To tackle the challenge of representativeness, we have previously created a study looking at the German PCSs [23, 24] where a large share of the PCSs in the country was under observation for the three months before the Corona pandemic. Since the data used in the previous study was obtained from publicly accessible websites, no information about the amount of energy charged, the users, etc. was available and data quality was limited by the collection method.

## B. This paper in the context of literature

This paper solves the previously existing issues by building on an exhaustive dataset provided by industry. This allows for a first-of-its-kind study in terms of data depth, quality, and are covered. The provided data encompasses the years 2019 until 2021 and provides PCS status for several ten thousand PCSs (number varies due to ongoing infrastructure expansion). This study further utilizes approximately 9 million charge detail records (CDRs) to provide insights on the energy demand distribution across charging processes. This rich dataset allows us to make statements regarding infrastructure use, user behaviour, and PCS profitability – each in a representative manner for Germany, the leading electric vehicle market by number of vehicles sold in Europe.

## C. Structure of this document

At first, the results obtained are shown in a repeating structure: Sections are started with a brief description of the topic followed by a list. The list is made up of headlines containing the key results and text blocks with details. The following section contains a discussion and conclusions drawn. The paper is terminated with a chapter describing the data and methodology used. This includes a brief discussion of the statistical validity of all shown plots. Results are shown before the extensive methodology discussion since all result chapters simply show measured data that is aggregated using averaging as well as simple estimated in section II.E.

## II. RESULTS

Results are shown in five categories. First, the energy consumption per charge event is discussed. In the next three sections, the CS usage is discussed in terms of when people arrive (II.B), how CSs are occupied at each moment in time (II.C), and how long CSs are occupied per charge event (II.D). As a last step, the profitability is estimated for German CSs.

## A. Energy consumption

Figure 3 shows the average power and energy consumption of charging processes split by rated power of the PCS. Figure 4 augments these findings by also showing the power levels achieved relative to rated power and split by connector type. Various aspects become apparent immediately:

### 1. A high rated power correlates with a high realized power

The higher the rated power, the higher is the average realized power in Figure 3 (a and b) for all power levels up until 200 kW and charging durations up to 8 hours (noise in the data increases afterwards). A difference can be seen comparing CHAdeMO and CCS in Figure 4. CHAdeMO PCSs achieve a lower power utilization.

The difference between the two connectors could be due to CHAdeMO only being installed on less than 4% of new vehicles[1]. CHAdeMO vehicles are consequently older on average and have a lower rated power (since rated power has gone up with time). The fact that no difference can be observed for chargers above and below 200 kW can be attributed to the fact that vehicles mostly cannot charge at power levels above ~150 kW.

---

[1] Own calculation based on [2, 25].

2. **Rated energy capacity remains largely unused**

   Across all connector types and rated powers, the average realized energy is between 30% and 60% of what the rated power would allow even for the charging processes with a very short duration (see Figure 4). The median value shown in Figure 10.c) and e) at 10 kWh after two hours correspond to ~25% and ~50% of the rated capacity for chargers rated at 12 – 25 kW and 4 – 12 kW respectively.

   For long sessions, this may be a result of batteries filling. For short sessions, vehicle charging power may be insufficient, e.g. due to insufficient DC capacity or not using all phases in AC charging [26].

3. **Start-up processes reduce realized power**

   Charge events of 15 minutes show lower power values compared to slightly longer charge events as is evident in Figure 3.a and Figure 4.

   The charging processes recorded in our database are typically started upon authentication (i.e. holding the charging card to the terminal or using apps or websites). Afterwards, the cable needs to be plugged in, a power negotiated between vehicle and PCS and the battery brought to a temperature range to accept the planned power flow. These steps possibly explain the low average power.

4. **After 4 - 5 hours, the realized energy does not increase at AC chargers**

   After some time, the realized energy at three-phase AC chargers is capped. On average this occurs after 4 hours resulting in a realized energy of 16.3 to 20.3 kWh (see orange and green line in Figure 3.b respectively). Figure 10 complements this finding since it can be seen that after 5 hours, 90% of charge events see no further increase in realized energy for the power levels between 4 and 25 kW.

5. **Fast-chargers with more than 100 kW rated power recharge 40 kWh in ~1 hour on average**

   Typical current batteries for long-range cars have a battery energy capacity between 60 and 100 kWh [27]. The average energy consumed reaches ~40 kWh as shown in Figure 3.b for EVSEs with more than 100 kW rated power.

   We can assume that this corresponds to a full recharge if some buffer is accounted for (see Figure 3.b). Median values shown in Figure 10 (see appendix) are similar.

6. **Less than 10% of Type 2 charge events realize the rated power**

   For Type 2 chargers below 25 kW, less than 10% of the shorter charge events make use of the rated charging power (see Figure 10). In addition, less than 50% of charge events at chargers with 22 kW consume 10.7 kWh over 2 hours.

Instead of aggregating by event duration, it is also possible to aggregate charge events by starting time (see Figure 5). The information to be gained from this figure can be summarized as follows:

1. **The energy consumed during AC charge events varies with the time of day**

   Charge events at AC PCSs create a high energy demand if started either in the evening or morning as the blue, orange, and green lines in Figure 5.a show. A similar pattern can be observed in the right column of Figure 10 in the top three plots.

   As will be outlined in section "Duration", this can be explained by the fact that events are significantly longer in the night or if started in the morning.

2. **Recharged energy per charge event has not changed significantly over time**

   The amount of energy delivered per charge event has not changed significantly over the observed period. While there are small changes in the average (Figure 5.b) and the median (appended dataset, not shown) of realized energy per charge event over time, the differences do not show a clear and consistent trend over time.

   While this finding is true for individual charge events, the increase of events shown in II.B means that the overall realized energy increased.

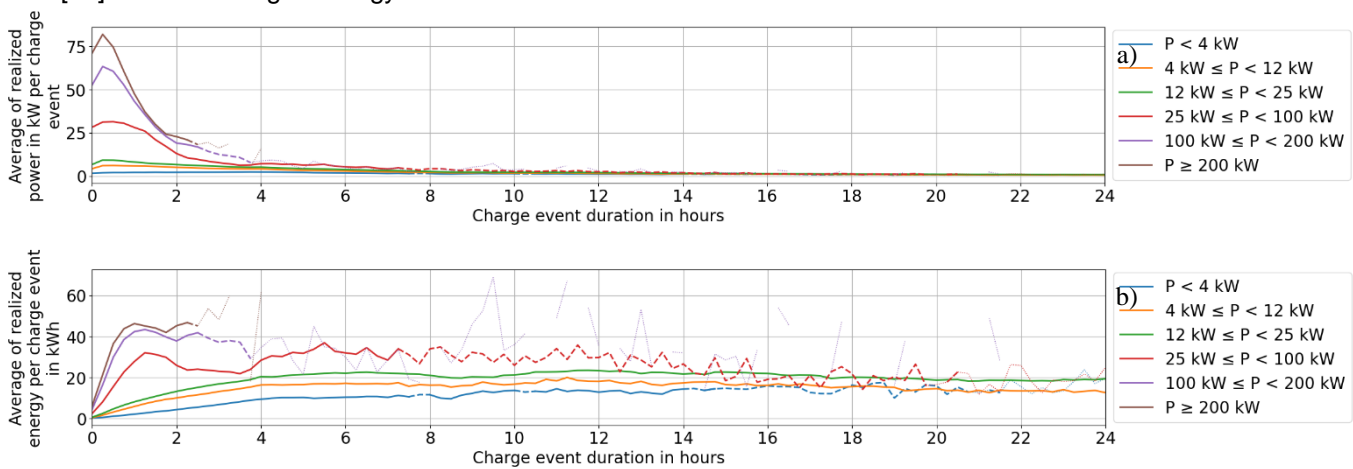

*Figure 3: Average actual power flow (a) and energy consumed (b) for the charge events recorded in the CDR data. Note that the lower graph can be generated by multiplying the upper graph with the x-axis value. Example on how to read the graphs: For charge events lasting around 30 minutes (with quarterhourly rounding windows) at chargers with a power rating above 200 kW,*

*the actual power consumption was 57 kW when averaging over the entire event duration. Dashed and dotted lines indicate a lower data quality as outlined in section XI.B in the appendix.*

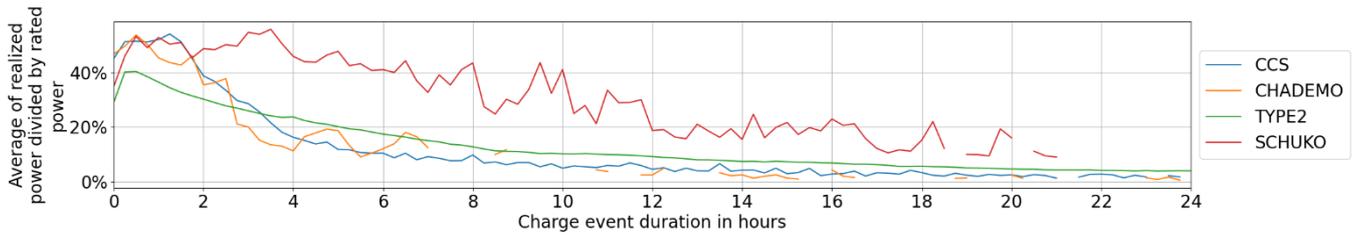

*Figure 4: Power and energy charged in a charge event relative to what would have been possible at rated power.*

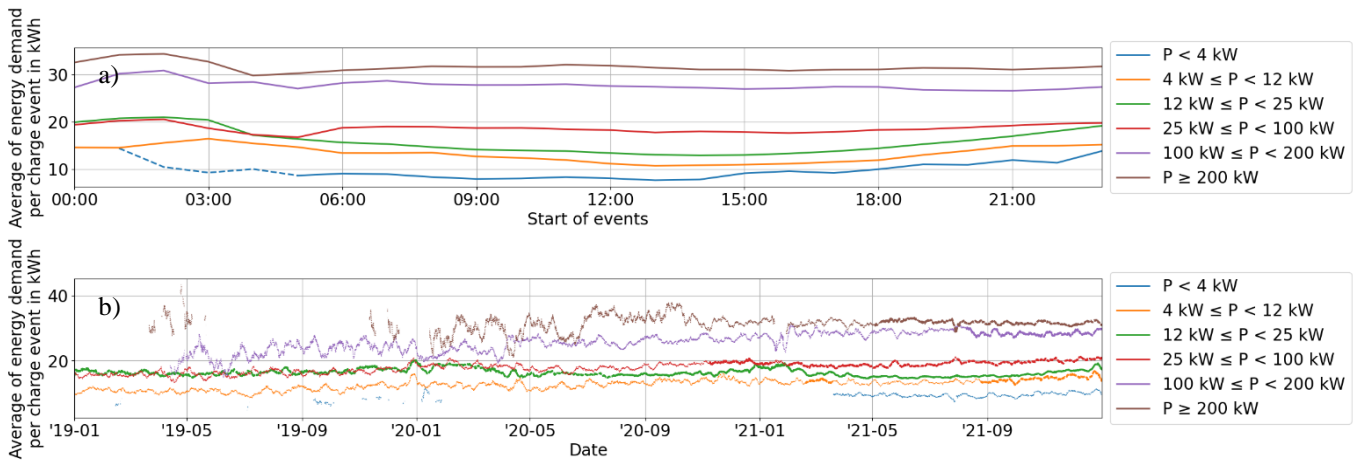

*Figure 5: Energy demand per charge event by time of arrival at the PCS. Example how to read a): During charge events starting at 3 PM at EVSEs with a rated power between 12 and 25 kW, 12.9 kWh are consumed on average. Values in b) are smoothed with a moving average of 1 week. Dashed and dotted lines indicate a lower data quality as outlined in XI.B in the appendix.*

*B. Arrivals*

Figure 6 shows the arrivals of vehicles at PCSs split by the selection criteria employed in this paper. Some key observations are:

1. **Number of arrivals per PCS have increased significantly over time**

   While the number of charge events increased for all PCS types over the observed period, the increase is stronger for EVSEs rated >100 kW (Figure 6.a).

2. **Changes in mobility patterns due to the Corona pandemic correlate with PCS usage**

   The Corona pandemic resulted in several lockdowns. The strongest reduction in mobility during the first lockdown (22 March – 4 May 2020) is reflected in decreasing charge events in Figure 6.a. Similar, but less dramatic effects, are visible for the subsequent lockdowns late 2020 and during the first half of 2021. The lower drops in usage correlate with a lower drop in mobility [28].

3. **The number of arrivals rises with increasing EVSE power**

   PCSs with a higher rated power experience more charge events than those with a low rated power (see Figure 6.b).

   This is can be linked to the lower average charging duration for each single EVSE.

4. **CCS is used significantly more than CHAdeMO**

   CHAdeMO connectors experience only about half as many charge events as CCS (Figure 6.c).

   A possible explanation is that CCS is the standard for fast-charging in Europe used by with virtually all European and US-American vehicles (including Tesla). EVSEs are however frequently equipped with both connectors [3] meaning that there are many more connectors per vehicle for CHAdeMO.

5. **Fast-charging is popular from Friday to Sunday while AC charging is reduced on Sundays**

   Fast-chargers experience on average more charge events on weekends as shown by the purple and brown line in Figure 6.b. For Type 2, the effect is opposite with lower usage on Sundays (Figure 6.c).

   For fast-chargers, this is presumably related to people taking longer trips on weekends than on weekdays where commuting is more common [29]. For Type 2 chargers, the lower number of events is likely the result of people going neither to work nor to other leisure activities such as shopping.

6. **The amount of arrivals follows a bell-shape with small deviations for CCS and Type 2**

   Most arrivals are almost symmetrical around noon with a slight increase between 6 and 9 AM for Type 2 connectors (Figure 6.c), particularly in industrial areas (Figure 6.d). For CCS, the bell-shape is slightly dented at midday with the two peaks at around 9 – 10 AM and 3 – 5 PM during workdays.

This is an indication of people charging their car while commuting. Events at industrial Type 2 connectors are frequently started in the morning. Weekend usage is characterised by the lack of a dent for fast-chargers (Figure 6.c).

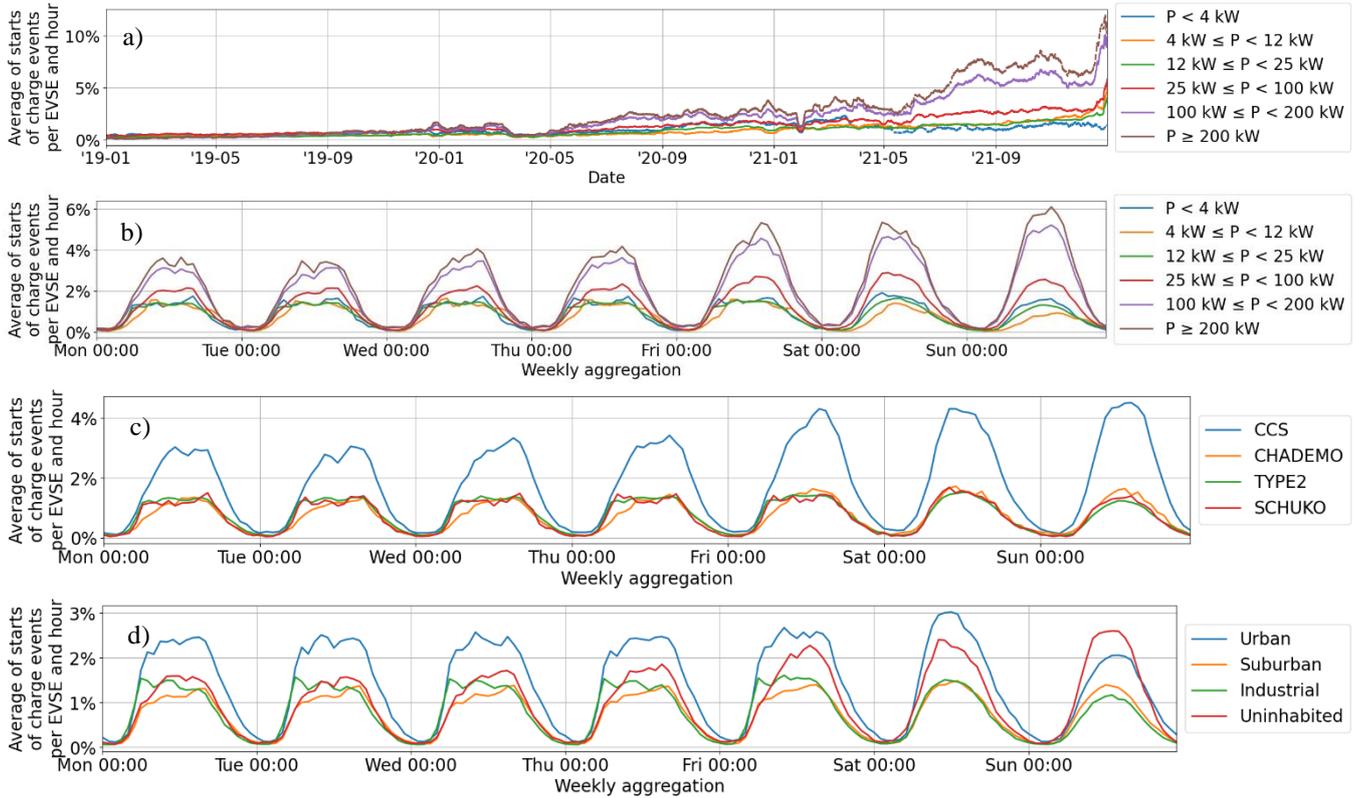

*Figure 6: Number of charge events by starting time of events split by power level over the observed period (a), aggregated over a week (b), split by connector type and aggregated over a week (c), and split by area type aggregated over a week (d). Example how to read: On an average Monday, at 3.0% of PCSs with a CCS connector a charge event is started between 2:30 and 3:30 PM (i.e. the hour around 3 PM). Additional aggregation levels as well as the same results for end of charge events are not shown here for the sake of conciseness, but are available in the appended dataset. Values in a) are smoothed with a moving average of 1 week.*

*C. Occupation*

Figure 7 shows the share of vehicles plugged in. Several trends are observable:

**1. Occupation is lower for fast-chargers**

Fast-chargers are occupied less time than their slower counterparts are (Figure 7.a and b).

This might be surprising given the fact that section II.B showed many more arrivals and departures for EVSEs with higher power ratings. The explanation for this phenomenon is that fast-charge events are much shorter than slow-charge events and therefore can service more vehicles while simultaneously also being much more available.

**2. Occupation is higher during the day**

The highest PCS occupation rate occurs during the day with the ratio between highest and lowest occupation rate being 1.4 – 1.8 across all power levels (Figure 7.b).

**3. Weekday-weekend patterns, CCS vs CHAdeMO, and long-running trends are similar to what can be observed for arrivals and departures**

The occupation follows the trends already visible in the number of arrivals and departures. The detailed explanations are consequently not repeated here (see section II.B).

**4. Standard deviation is large**

The standard deviation (appended dataset, not shown) of all lines in Figure 7 is larger than the average. Typical values are between 20 and 40%.

This observation indicates a large diversity and randomness in the underlying dataset. As discussed in section IV.E, this was to be expected given the low predictive value of the power level, connector type and surrounding area type alone [30]. Individual circumstances must therefore be kept in mind when planning and operating a PCS.

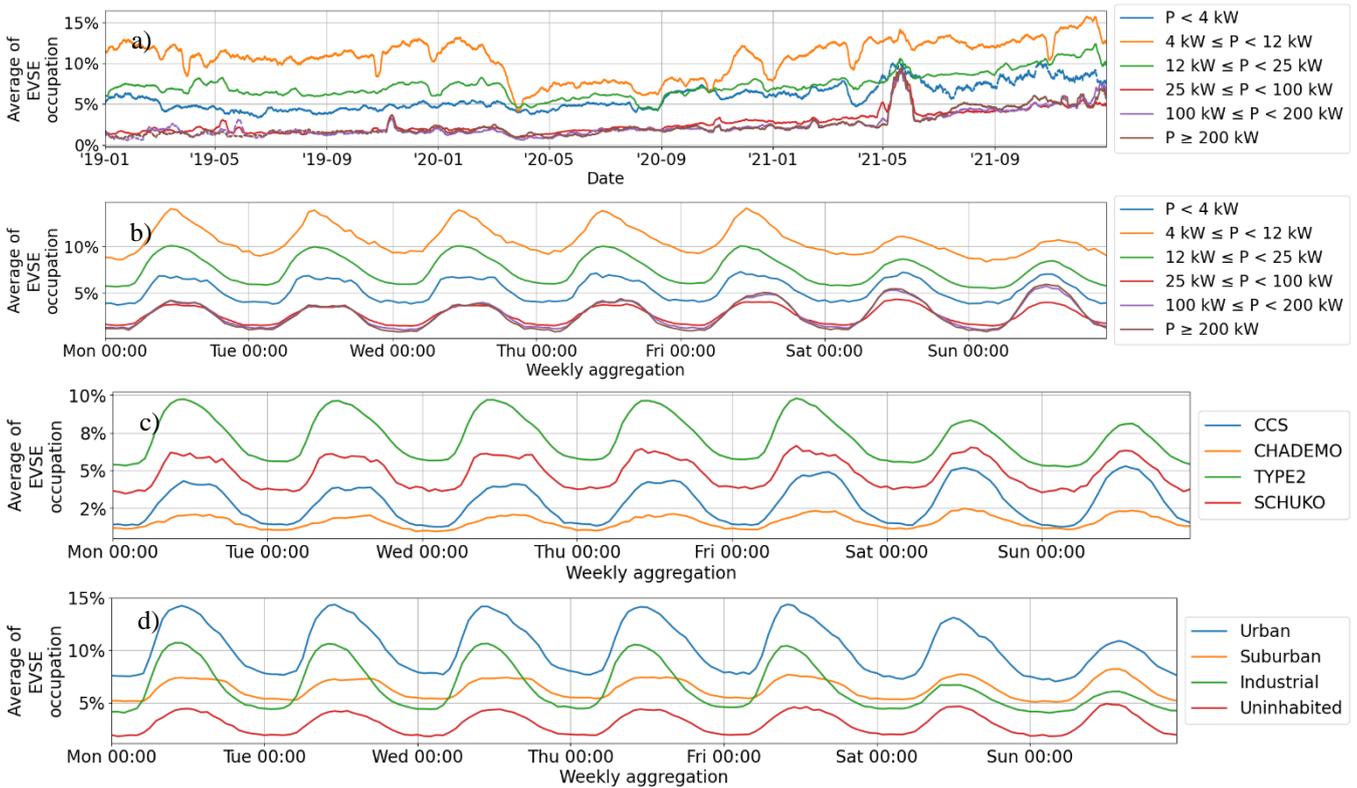

Figure 7: Share of PCSs occupied split by power level over the observed period (a), aggregated over a week (b), split by connector type and aggregated over a week (c), and split by area type aggregated over a week (d). Example how to read: On an average Monday at 3 PM, 9.85% of PCSs with a power rating between 12 and 25 kW were occupied. Values in a) are a one-week moving average. Dashed and dotted lines indicate a lower data quality as outlined in XI.B in the appendix.

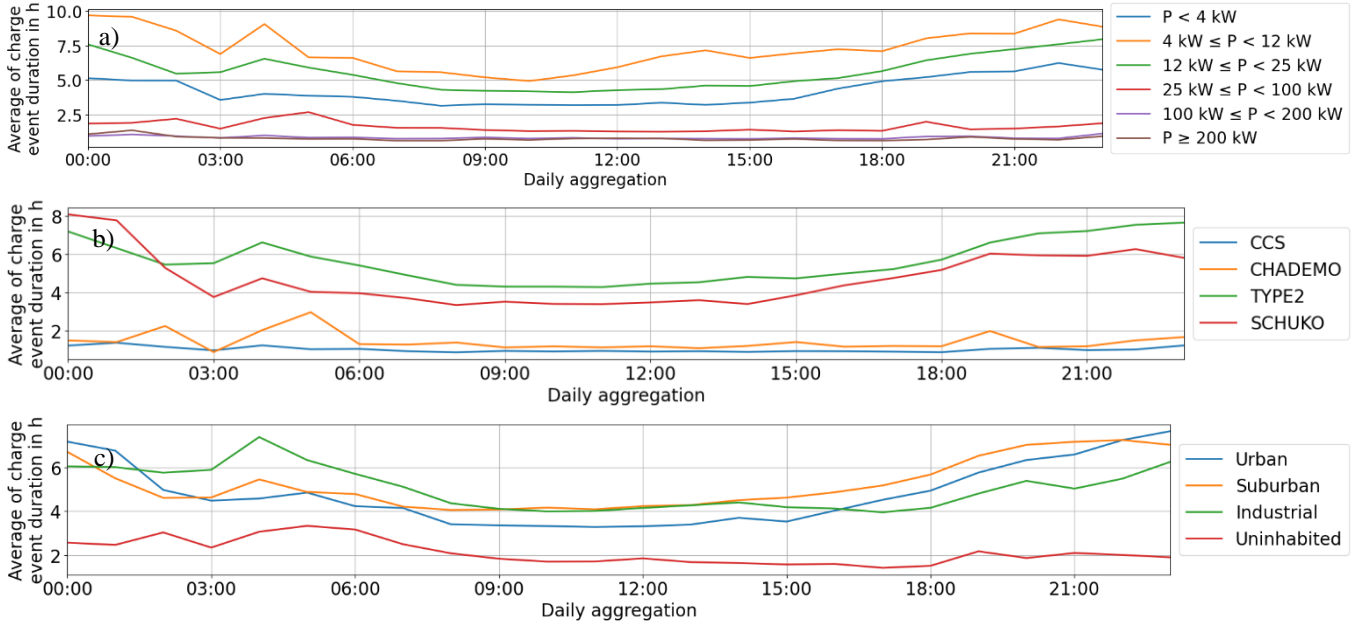

Figure 8: Plot of the duration of charge events by the time that they started aggregated by week and split by rated EVSE power (a), connector type (b) and land usage (c)

### D. Duration

Figure 8 shows the duration of charging processes where the value is recorded for the time step when the charge event was started. The key findings are:

#### 1. AC charge events are longer if they start in the morning or evening than at noon

Charge events at AC CSs are longer if starting between 7 PM and 9 AM (see Type 2 and Schuko in Figure 9.b). The effect is stronger for median values (appended data, not shown) where the median duration at noon is 2 hours for all AC-chargers compared to 5.6 and 8.8 hours at midnight for connectors rated 12 – 25 kW and 4 – 11 kW respectively.

This is likely the combination of people parking their vehicles during either the night or workday.

2. **Events at industrial sites are on average especially long if started between 3 and 9 AM**

   As Figure 8.c shows, charge events in industrial areas are long if they are started between 3 and 9 AM. The pattern is inverted in urban settings, which experience long occupations during the night.

   This can be linked to people leaving their vehicle parked for the workday in industrial areas for the whole day. This is not possible in urban areas where parking is typically limited to 2 – 4 hours [31].

*E. Profitability*

Figure 9 shows an estimation of the annual energy sales per EVSE sorted by increasing sales. Table 1 additionally shows a sample calculation of required energy sales. Based on assumptions regarding sales margin, costs, interest rate and lifetime, the share of EVSEs is calculated that covers their costs through energy sales. Typical sales margin in the authors' industry experience are marked yellow.

1. **Energy sales per year are nearly identical for all fast-chargers with a power rating above 100 kW**

   There is virtually no difference in energy sales at fast-PCSs rated above 100 kW.

   This likely because many vehicles cannot use these power levels yet.

2. **20% of AC EVSEs sell ~55% of the energy**

   The highly inhomogeneous pattern in Figure 9 means that 20% of EVSEs with a power rating between 4 and 100 kW sell approximately 55% of the energy in that group. For high power DC chargers, this value is reduced to 40%.

3. **Vast majority of EVSEs is unprofitable**

   In Table 1 PCSs with a high power rating appear most profitable with 18% to 61% of EVSEs achieving break-even by the sale of electricity. For slower chargers, the situation is more problematic. Although investment costs are a magnitude lower, electricity sales are also equally lower. These CSs compete with household prices of ~30 €ct/kWh while purchasing electricity at ~20 €ct/kWh [32].

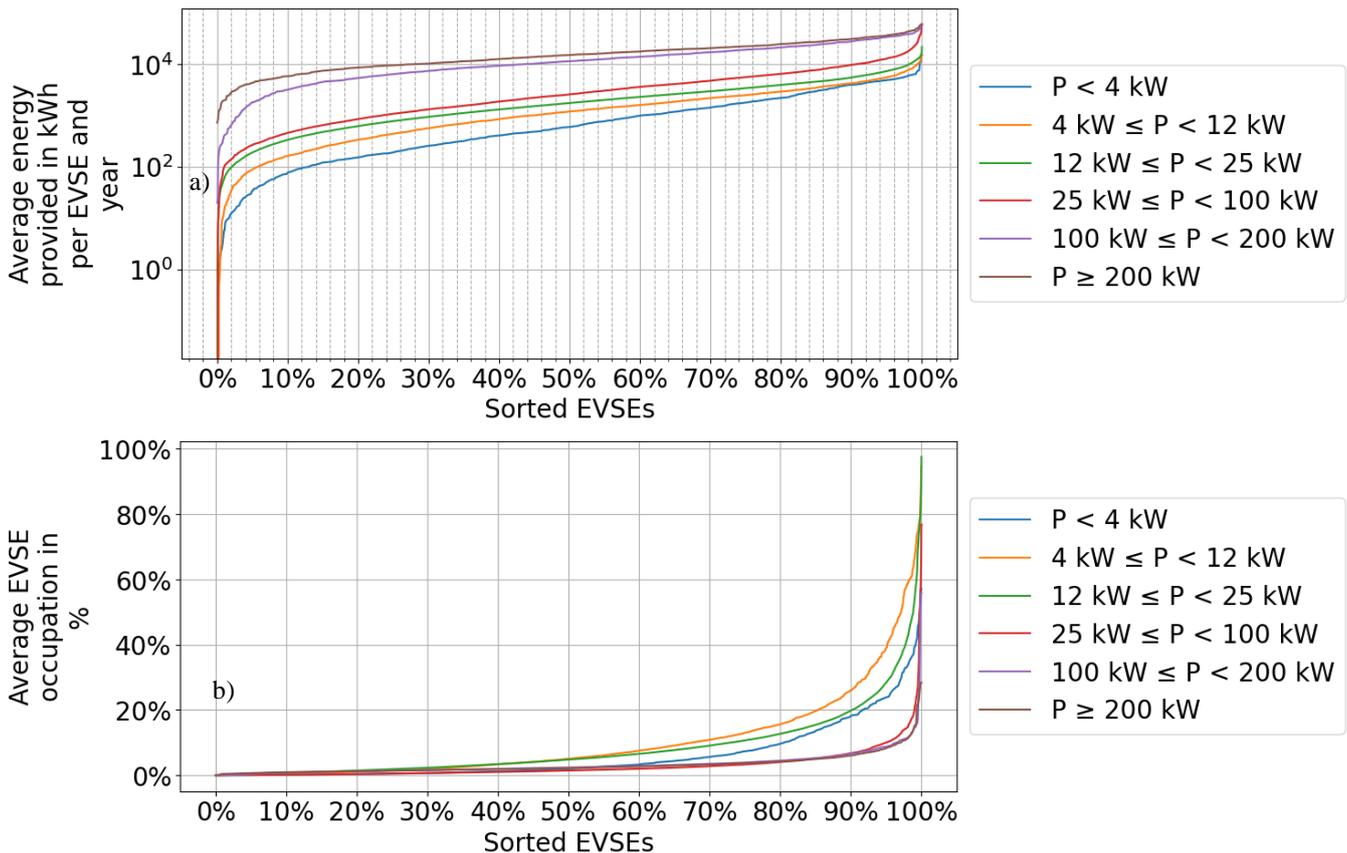

Figure 9: Average energy sold (a) and average occupation (b) per EVSE categorized by power level. Example how to read: 10% of PCSs with a rated power between 12 and 25 kW sell 336 kWh or more per year.

Table 1: Estimation of share of profitable EVSEs given the assumptions made in the first five rows in the table with OpEx and interest rate being quite optimistic. Cells marked in yellow are most realistic sales margins according to the authors' experience, but do not include the potential revenues from greenhouse gas emission quota trading. The first block of results is the aggregate of all area types and the subsequent blocks are split by area type. Note that costs were kept constant across different area types to ensure comparability which may conflict with reality.

| | | P < 4 kW | 4 kW ≤ P < 12 kW | 12 kW ≤ P < 25 kW | 25 kW ≤ P < 100 kW | 100 kW ≤ P < 200 kW | P ≥ 200 kW |
|---|---|---|---|---|---|---|---|
| | Investment cost | 2,000 € | 6,000 € | 9,000 € | 50,000 € | 100,000 € | 200,000 € |
| | Lifetime in years | 8 | 8 | 8 | 8 | 8 | 8 |
| | Real interest rate | 4% | 4% | 4% | 4% | 4% | 4% |
| | OpEx in €/a | 122.2 | 76.3 | 77.2 | 136.4 | 128.8 | 117.6 |
| | Number of EVSEs per PCS | 3.94 | 2.41 | 2.44 | 3.31 | 3.12 | 2.84 |
| Share profitable by sales margin in €ct/kWh | | PCS independent of area type | | | | | |
| | 5 | 21.29% | 1.77% | 1.04% | 0.08% | 0.08% | 0.11% |
| | 10 | 37.72% | 11.63% | 9.09% | 1.04% | 0.42% | 0.11% |
| | 15 | 47.23% | 23.15% | 20.88% | 3.07% | 6.23% | 0.11% |
| | 20 | 52.83% | 33.04% | 31.12% | 6.93% | 14.99% | 0.47% |
| | 30 | 63.12% | 46.51% | 46.62% | 15.57% | 33.13% | 6.26% |
| | 40 | 68.99% | 55.39% | 56.81% | 24.46% | 47.77% | 17.53% |
| | 50 | 73.50% | 61.40% | 64.05% | 31.60% | 58.92% | 29.10% |
| | 60 | 77.06% | 66.47% | 69.45% | 38.20% | 67.34% | 40.79% |
| | | PCSs in an urban environment | | | | | |
| | 5 | 50.02% | 7.73% | 3.87% | 0.39% | 2.94% | No data |
| | 10 | 66.77% | 34.55% | 24.72% | 3.11% | 2.94% | |
| | 15 | 75.29% | 48.92% | 42.67% | 10.59% | 4.17% | |
| | 20 | 76.29% | 58.43% | 54.84% | 19.11% | 11.92% | |
| | 30 | 80.74% | 67.76% | 69.16% | 35.24% | 15.88% | |
| | 40 | 84.31% | 73.54% | 76.67% | 44.32% | 34.80% | |
| | 50 | 84.79% | 78.64% | 81.73% | 53.71% | 53.07% | |
| | 60 | 85.88% | 80.62% | 85.11% | 56.76% | 60.18% | |
| | | PCSs in a suburban environment | | | | | |
| | 5 | 14.79% | 1.14% | 0.56% | 0.20% | 0.40% | 1.21% |
| | 10 | 31.18% | 12.19% | 7.07% | 0.65% | 0.40% | 1.21% |
| | 15 | 40.54% | 24.84% | 18.55% | 1.86% | 2.20% | 1.21% |
| | 20 | 47.48% | 36.38% | 29.04% | 5.06% | 5.23% | 1.21% |
| | 30 | 59.49% | 50.47% | 45.13% | 12.57% | 17.95% | 6.06% |
| | 40 | 66.75% | 60.95% | 55.77% | 21.45% | 28.77% | 10.55% |
| | 50 | 71.76% | 65.75% | 63.19% | 27.57% | 37.31% | 21.11% |
| | 60 | 76.06% | 70.83% | 68.86% | 34.27% | 45.95% | 31.37% |
| | | PCSs in an industrial environment | | | | | |
| | 5 | 19.74% | 0.74% | 0.69% | 0.11% | 0.25% | 0.40% |
| | 10 | 37.26% | 5.79% | 6.08% | 1.43% | 0.67% | 0.40% |
| | 15 | 48.64% | 16.75% | 16.89% | 4.06% | 7.26% | 0.40% |
| | 20 | 54.96% | 26.34% | 26.84% | 8.70% | 18.25% | 1.22% |
| | 30 | 64.97% | 41.51% | 43.23% | 18.79% | 33.36% | 7.70% |
| | 40 | 68.24% | 49.93% | 54.02% | 28.08% | 45.62% | 22.22% |
| | 50 | 73.46% | 56.83% | 61.59% | 36.52% | 55.74% | 36.31% |
| | 60 | 75.95% | 62.41% | 67.18% | 43.94% | 63.71% | 45.47% |
| | | PCSs in an uninhabited environment | | | | | |
| | 5 | 5.00% | 0.73% | 0.16% | 0.08% | 0.17% | 0.18% |
| | 10 | 12.85% | 1.36% | 2.50% | 0.62% | 0.58% | 0.18% |
| | 15 | 16.78% | 4.78% | 7.02% | 1.53% | 7.67% | 0.18% |
| | 20 | 18.45% | 8.07% | 12.06% | 3.95% | 17.42% | 0.34% |
| | 30 | 20.63% | 16.35% | 22.21% | 10.17% | 40.91% | 6.06% |
| | 40 | 24.75% | 24.95% | 32.31% | 18.47% | 58.57% | 17.06% |
| | 50 | 40.21% | 34.26% | 41.71% | 24.41% | 71.04% | 27.65% |
| | 60 | 51.59% | 41.89% | 48.34% | 30.56% | 79.96% | 40.94% |

## III. DISCUSSION AND CONCLUSION

The results shown in this paper provide detailed insights into the way that PCSs in Germany are used. The value of the work performed lies in the ability to quantify behaviours and effects that previously had an anecdotal character. An example of this type of result is that it was suspected that AC PCS are frequently used for parking without much energy flowing. Figure 10.e quantifies this intuition by showing that median AC charge events at 11 kW PCSs transfer only 11.75 kWh over a 4 h period compared to a theoretical maximum of 44 kWh. EVSEs with more than 200 kW rated power on the other hand show a median energy transfer of 46 kWh after 1 h, which is likely much closer to realistic values for fully charging a vehicle. Users could therefore use 11 kW PCS only 27% of the time if they actually fully recharged their vehicle at every charge event. Other observations of similar nature are possible using the provided material.

Charge point operators, public planners, mobility experts, scholars in the fields of mobility and energy, and other groups benefit from the obtained results. Applications of the found values include estimating the profitability of future PCS sites, comparing the performance of the current portfolio with national averages, comparing modelling results with real-world data, understanding customer behaviour, and many others.

Some limitations must be considered when working with the results. As for instance Figure 6.a shows, the usage patterns of PCSs are far from constant over time. Reasons are among others the effect of the Corona pandemic on human mobility as well as the exponential increase in electric vehicle stock. When working with the average weekly usage data, readers should consequently ensure that the data used fits their assumptions regarding human mobility. Since there have been such a large variety of situations, we decided to not display each possible combination of market update and corona counter-measures. Instead, we invite scholars, who want to work with the data, to use the non-aggregated data together with the number of observations in the attached dataset.

When combining the results of this paper, many conclusions can be drawn and only a short list is provided here for the sake of conciseness.

1. Most PCSs are not profitable from the energy that they sell according to our estimation shown in Table 1. This is particularly true for the slower AC chargers (38%, 12%, and 9% for PCS in the power ranges <4 kW, 4 – 12 kW, and 12 – 25 kW respectively). The lack of profitability is assumingly a product of the lower margins compared to DC PCSs. This is comparable to gas stations' small margins on gas sales [33, 34]. PCSs should therefore be combined with the more profitable in-store sales and other secondary services.

2. There are strong differences between the PCSs in each category where the best-performing fast-charging EVSE with a power rating above 100 kW is estimated to deliver 61 MWh per year and the worst-performing only about 20 kWh.

3. Clear day-night and weekday-weekend patterns can be seen. AC chargers are used predominantly during the week and less on Saturdays. For the typical 22 kW chargers, the highest average occupation on weekdays is for instance 10% while it is only 8% on weekends. In industrial settings, this effect is enlarged by the high relevance of commuters resulting in a weekday peak occupation of 11% and only 7% on weekends. Fast-chargers are used mostly on weekends. Starting Friday afternoon, charge events are much more frequent during times when people take long-distance trips. This results in the average occupation being 3% to 3.5% higher on weekends compared to weekdays.

4. Dynamic pricing and other incentive mechanisms should be implemented at PCSs to make sure that the low occupation of PCS during the night and other offpeak periods is avoided. Otherwise, the low load factor results in a challenging business case for the PCS operator. In the project BeNutz LaSA, prediction algorithms [30] and incentive strategies are created to achieve this goal.

## IV. METHODS

### A. Data sources

This section lists the data sources that were used in the context of this document. These comprise of the PCS as the main data source. This data is split into the more precise charge detail records (CDRs), which hold information about individual charge events, and the PCS status data, listing when a PCS is available, occupied, etc. The second data source are area types which are based on the Corine land cover model of the European Union [35]. Other external data sources such as weather data, traffic information or public holidays were also available to the authors, but are not included in this paper since a previous study showed that they had limited explanatory value [30].

#### 1) PCS data

In the context of the project BeNutz LaSA [36], our industry partners Hubject and SMART/LAB provided us with usage data of 27,800 PCSs. Both datasets contain the location of each PCS, the ids and rated power of the connected EVSEs, the connectors available at each EVSE and the status changes of the EVSEs. To simplify the analysis and ensure consistency across the dataset, we limit our analysis to the statuses "occupied", "available", and "unknown". Other status reports such as "reserved", "preparing", etc. were mapped one of the three selected as outlined in [24].

*Table 2: Example of how the status changes for an EVSE would be translated into the various time series assuming an hourly resolution. "O" – EVSE became occupied, "A" – EVSE became available, and "U" that the status of the EVSE became unknown.*

| Timestep [h] | 1 | 2 | 3 | 4 | 5 | 6 | 7 | 8 | 9 | 10 | 11 | 12 | 13 | 14 | 15 | 16 | 17 | 18 | 19 | 20 | 21 | 22 | 23 | 24 |
|---|---|---|---|---|---|---|---|---|---|---|---|---|---|---|---|---|---|---|---|---|---|---|---|---|
| Raw data | | O | | | A | | U | | A | | | | | O | | | | A | | O | | U | O | A |
| Starts [] | 0 | 1 | 0 | 0 | 0 | 0 | 0 | 0 | 0 | 0 | 0 | 0 | 0 | 1 | 0 | 0 | 0 | 0 | 0 | 1 | 0 | 0 | 0 | 0 |
| Duration [h] | | 3 | | | | | | | | | | | | 4 | | | | | | 4 | | | | |
| Energy [kWh] | | 15 | | | | | | | | | | | | 24 | | | | | | 24 | | | | |
| Occupation [] | 0 | 1 | 1 | 1 | 0 | 0 | 0 | 0 | 0 | 0 | 0 | 0 | 0 | 1 | 1 | 1 | 1 | 0 | 0 | 1 | 1 | 1 | 1 | 0 |

A second key piece of data are the CDRs provided by the industry partners. Each such record contains details about a charge event such as the start and end time, the energy that was consumed during the event, and the id of the used EVSE. For one of the partners, the records further contained an anonymized id of the contract and the authentication token used.

*2) Area types*

The Corine land cover model [35] was used to identify the area type of PCSs. The model assigns a land usage type to land across the European Union. The available land use categories were grouped into the categories Urban, Suburban, Industrial, Uninhabited and Non-Fitting (for details see [24]).

### B. Data cleaning

*1) Limiting time window*

Since several results shown rely on knowing the durations of charge events, it was necessary to discard all data from the last status change of PCSs onwards as the duration of the last status would be undefined.

*2) Removal of PCSs with little status information*

To generate useful time series data, EVSEs were only considered if they experienced at least two status changes with the valid timeframe being between the first and the last status change. This was done to prevent unrealistically long timespans if PCSs stop reporting data.

*3) Removal of PCSs from outside Germany*

Some of the PCSs obtained in the original dataset were just outside of the German borders. We removed all PCSs that were not located within the boundaries of Germany as defined by [37].

*4) Removal of CDRs with unrealistic values*

Unrealistic CDRs were removed from the dataset if their consumed energy was either equal to 0 or above 1 MWh. They were also removed if information about their rated power was unavailable or invalid. The valid ranges were selected as shown in Table 3.

*Table 3: Valid combinations of connector types and power and voltage ranges. For Type 2 and Schuko, we converted values to phase-to-neutral values.*

| Connector | Valid rated power range | Valid voltage range |
|---|---|---|
| Type 2 | $10\ kW \leq P \leq 43\ kW$ | $U \geq 220V$ |
| Schuko | $P \leq 4\ kW$ | $U \geq 220V$ |
| CCS, CHAdeMO | $P \geq 50\ kW$ | $U \geq 400V$ |

### C. Time series generation

The data provided by the industry partners contains only status changes and not an entire time series. The events therefore had to be translated into time series data. This was done in the following steps for each EVSE in a slightly different way depending on the calculated value (see also Table 2 for an example):

1. **Arrivals:** If an event was started in a time step, the time step was marked as 1, otherwise 0. If periods with an unknown status occurred, a 1 would only be assigned if the status after the unknown period was different from the status before.

2. **Occupation**: An EVSE was marked as 1 for all moments in time where the previous status was "occupied" and as 0 for all moments in time when the last reported status was "available".

3. **Duration**: Similar to "Starts", but the duration in hours instead of a 1 was used in the time step when the event started. For all duration plots, the durations shown consequently correspond to when events were started.

4. **Energy consumption**: Similar to "Duration", but the average energy consumption for events of the given duration was looked up in the CDRS.

### D. Profitability estimation

The profitability is calculated using the following steps.

CapEx is a single payment whereas OpEx and revenues are continuous cash flows. To allow comparison between the two, the CapEx are annualized using the annuity factor calculation below. Let $ANF$ be the annuity factor, $i$ the real interest rate defined as the nominal interest rate minus the inflation rate, $n$ the number of years, and $CapEx_{Ann}$ the annualized CapEx.

$$ANF = \frac{(1+i)^n \cdot i}{(1+i)^n - 1}$$
$$CapEx_{Ann} = ANF \cdot CapEx$$

To determine the amount of energy that a PCS would need to sell in order to cover its costs, the following steps were taken:

1. $CapEx_{Ann}$ was added to OpEx to calculate the full annual recurring costs $RC$:

$$RC = CapEx_{Ann} + OpEx$$

2. The amount of energy required is found by dividing $RC$ by the sales margin per unit of energy $\lambda$ as follows:

$$E_{required} = \frac{RC}{\lambda}$$

### E. Statistical validity

The values in this paper are representative in the sense that they are based on a large dataset comprising the majority of PCSs in the country. We, nevertheless, made several choices with regards to visualization of results and analysis which are justified in the following.

*1) Choice of features*

The major features that this paper analyses are power level, location of the PCS, and connector type. These features were chosen based on previous analysis [24, 38] where these showed consistently low p-values below 0.05 and comparatively high predictive power [30]. These two aspects combined indicate that there is a strong and non-random correlation between the chosen features and the displayed results.

Other features such as the weather, vacation periods and long weekends are not considered. The reason for not including the weather is that neither temperature nor precipitation had strong predictive quality [30]. All three features are further highly seasonal. Since the displayed data encompasses the Corona pandemic, such a seasonality might be misleading since there were much stronger mobility restrictions in place during the lock-downs during the winters as compared to summers [28]. Any observed effect may therefore be correlated to the unusual circumstances caused by the pandemic and not necessarily be valid in the future anymore.

*2) Average values*

The results reported in this document are all presented in terms of average values since this comparatively simple measure can be applied across all reported results and is easily understood in the wider scientific community. The main problem with this approach is the fact that the data is typically not normally distributed around the average as Figure 10 shows. This leads to the averages being substantially higher than the median.

Other measures such as plotting deciles (i.e. 10%-steps) or medians could arguably be used instead at least in some visualisations. For the sake of conciseness, we decided against showing these metrics in the paper. Instead, they are provided in the attached dataset for readers wishing to perform a deep-dive into the data.

*3) Standard deviations and uncertainty*

As stated in section IV.E.1), the features were chosen based on p-values and predictive power. The usage patterns of PCSs are, however, subject to a large number of (random) factors and consequently cannot be explained using a few simple features alone. One way to measure this effect is using the standard deviation around the average. As outlined in the previous section, the data is not normally distributed around the average, which would normally not allow for the usage of standard deviation. Since the metric is again one understood by the wider community, we nevertheless calculated it for all displayed plots and added it to the additional material. The second measure is the decile plot, which provides a more detailed insight in the data at the expense of requiring one plot per line a line-plot.

## V. ACKNOWLEDGEMENTS

We would like to thank our industry partners at SMART/LAB and Hubject for the provision of the data upon which the presented research is largely based and for the support when questions regarding the data arose. We would further like to thank the team from the institutes ISEA, IAEW and FCN (all RWTH Aachen University) and Lennart Jansen for his helpful comments.

## VI. FUNDING INFORMATION

The research, on which this paper is based, was done in the context of the project BeNutz LaSA. This Project is supported by the Federal Ministry for Economic Affairs and Climate Action (BMWK) on the basis of a decision by the German Bundestag. The funding identifier is "01MV20001A".

## VII. AUTHOR CONTRIBUTION STATEMENT

Author contributions are given according to the CRediT framework as follows: Christopher Hecht – conceptualization, methodology, software, validation, formal analysis, investigation, data curation, writing of the original draft, visualization, funding acquisition; Jan Figgener – conceptualization, validation, funding acquisition, supervision, review and editing, project administration; Dirk Uwe Sauer – Resources, review and editing, supervision, project administration, funding acquisition.

## VIII. ADDITIONAL MATERIAL

This preprint is provided without additional material. The paper was submitted to the journal iScience and is currently undergoing review. The additional material will be published jointly with the journal paper.

## IX. GLOSSARY

| Term | Definition |
| --- | --- |
| Rated power/energy | The maximum power/energy that a system is able to provide<br>Synonyms: installed, available |
| Realized power/energy | The actual power/energy that flew during a charge event |
| CCS | "Combined Charging System". A fast-charging connector standard |
| CHAdeMO | "CHArge de MOve". A fast-charging connector standard |
| Type 2 | The three-phases connector standard used in Europe |
| Schuko | Single-phase plug and socket standard that is also used for normal household applications |
| Fast-charging | Refers to charging with high power. Traditionally, power ratings of 43 kW and more were considered. This paper defines fast-charging starting at 100 kW due to the otherwise long charging durations. |
| Charge event | The chain of events starting with a user authenticating at a PCS and ending with the unplugging of the vehicle. |
| AC | "Alternating Current" |
| DC | "Direct Current" |
| EVSE | "Electric Vehicle Supply Equipment"<br>The device powering one or many connectors out of which only one can be used at a time (e.g. if CHAdeMO and CCS are installed at the same EVSE and only one can be used) |

| Connector | The physical connection point on the vehicle or PCS. The shape is defined by the connector standard. |
| PCS | "Charging Station" A collection of EVSEs |
| CDR | "Charge Detail Record" A data entry containing details about a charge event such as the start and end time, energy transferred and EVSE used. |

X. REFERENCES

[1] CAIT Climate Data Explorer via. Climate Watch, *Emissions by sector*. [Online]. Available: https://ourworldindata.org/emissions-by-sector#annual-co2-emissions-by-sector (accessed: Jan. 27 2022).

[2] Kraftfahrt-Bundesamt, *Bestand am 1. Januar 2021 nach Zulassungsbezirken und Gemeinden*. [Online]. Available: https://www.kba.de/DE/Statistik/Fahrzeuge/Bestand/ZulassungsbezirkeGemeinden/zulassungsbezirke_node.html (accessed: Jan. 27 2022).

[3] Bundesnetzagentur, *Ladesäulenkarte*. [Online]. Available: https://www.bundesnetzagentur.de/DE/Sachgebiete/ElektrizitaetundGas/Unternehmen_Institutionen/HandelundVertrieb/Ladesaeulenkarte/Ladesaeulenkarte_node.html (accessed: Jan. 27 2022).

[4] A. Almaghrebi, S. Shom, F. Al Juheshi, K. James, and M. Alahmad, "Analysis of User Charging Behavior at Public Charging Stations," in *2019 IEEE Transportation Electrification Conference and Expo (ITEC)*, Detroit, MI, USA, 2019, pp. 1–6.

[5] A. Almaghrebi, F. A. Juheshi, J. Nekl, K. James, and M. Alahmad, "Analysis of Energy Consumption at Public Charging Stations, a Nebraska Case Study," in *2020 IEEE Transportation Electrification Conference & Expo (ITEC)*, Chicago, IL, USA, 2020, pp. 1–6.

[6] A. Almaghrebi, F. Aljuheshi, M. Rafaie, K. James, and M. Alahmad, "Data-Driven Charging Demand Prediction at Public Charging Stations Using Supervised Machine Learning Regression Methods," *Energies*, vol. 13, no. 16, p. 4231, 2020, doi: 10.3390/en13164231.

[7] R. van den Hoed, J. R. Helmus, R. de Vries, and D. Bardok, "Data analysis on the public charge infrastructure in the city of Amsterdam," in *2013 World Electric Vehicle Symposium and Exhibition (EVS27)*, Barcelona, Spain, 2013, pp. 1–10.

[8] R. Wolbertus, R. van den Hoed, and S. Maase, "Benchmarking Charging Infrastructure Utilization," *WEVJ*, vol. 8, no. 4, pp. 754–771, 2016, doi: 10.3390/wevj8040754.

[9] R. Wolbertus, M. Kroesen, R. van den Hoed, and C. Chorus, "Fully charged: An empirical study into the factors that influence connection times at EV-charging stations," *Energy Policy*, vol. 123, pp. 1–7, 2018, doi: 10.1016/j.enpol.2018.08.030.

[10] M. K. Gerritsma, T. A. AlSkaif, H. A. Fidder, and W. G. van Sark, "Flexibility of Electric Vehicle Demand: Analysis of Measured Charging Data and Simulation for the Future," *WEVJ*, vol. 10, no. 1, p. 14, 2019, doi: 10.3390/wevj10010014.

[11] M. van der Kam, W. van Sark, and F. Alkemade, "Multiple roads ahead: How charging behavior can guide charging infrastructure roll-out policy," *Transportation Research Part D: Transport and Environment*, vol. 85, p. 102452, 2020, doi: 10.1016/j.trd.2020.102452.

[12] Z. Zhang, Z. Chen, Q. Xing, Z. Ji, and T. Zhang, "Evaluation of the multi-dimensional growth potential of China's public charging facilities for electric vehicles through 2030," *Utilities Policy*, vol. 75, p. 101344, 2022, doi: 10.1016/j.jup.2022.101344.

[13] J. Betz, L. Walther, and M. Lienkamp, "Analysis of the charging infrastructure for battery electric vehicles in commercial companies," in *2017 IEEE Intelligent Vehicles Symposium (IV)*, Los Angeles, CA, USA, Jun. 2017 - Jun. 2017, pp. 1643–1649.

[14] D. Efthymiou, K. Chrysostomou, M. Morfoulaki, and G. Aifantopoulou, "Electric vehicles charging infrastructure location: a genetic algorithm approach," *Eur. Transp. Res. Rev.*, vol. 9, no. 2, p. 596, 2017, doi: 10.1007/s12544-017-0239-7.

[15] D. Ji *et al.*, "A Spatial-Temporal Model for Locating Electric Vehicle Charging Stations," in *Communications in Computer and Information Science, Embedded Systems Technology*, Y. Bi, G. Chen, Q. Deng, and Y. Wang, Eds., Singapore: Springer Singapore, 2018, pp. 89–102.

[16] M. Cocca, D. Giordano, M. Mellia, and L. Vassio, "Data Driven Optimization of Charging Station Placement for EV Free Floating Car Sharing," in *2018 21st International Conference on Intelligent Transportation Systems (ITSC)*, Maui, HI, Nov. 2018 - Nov. 2018, pp. 2490–2495.

[17] A. Pahlavanhoseini and M. S. Sepasian, "Scenario-based planning of fast charging stations considering network reconfiguration using cooperative coevolutionary approach," *Journal of Energy Storage*, vol. 23, pp. 544–557, 2019, doi: 10.1016/j.est.2019.04.024.

[18] T. S. Bryden, G. Hilton, A. Cruden, and T. Holton, "Electric vehicle fast charging station usage and power requirements," *Energy*, vol. 152, pp. 322–332, 2018, doi: 10.1016/j.energy.2018.03.149.

[19] M. Baresch and S. Moser, "Allocation of e-car charging: Assessing the utilization of charging infrastructures by location," *Transportation Research Part A: Policy and Practice*, vol. 124, pp. 388–395, 2019, doi: 10.1016/j.tra.2019.04.009.

[20] M. Erbaş, M. Kabak, E. Özceylan, and C. Çetinkaya, "Optimal siting of electric vehicle charging stations: A GIS-based fuzzy Multi-Criteria Decision Analysis," *Energy*, vol. 163, pp. 1017–1031, 2018, doi: 10.1016/j.energy.2018.08.140.

[21] M. Fischer, C. Hardt, W. Michalk, and K. Bogenberger, *Charging or Idling: Method for Quantifying the Charging and the Idle Time of Public Charging Stations: Conference: Transportation Research Board (TRB) 101st Annual Meeting in Washington D.C.* [Online]. Available: https://www.researchgate.net/publication/357900426_Charging_or_Idling_Method_for_Quantifying_the_

Charging_and_the_Idle_Time_of_Public_Charging_Stations (accessed: Feb. 8 2022).
[22] L. Calearo, M. Marinelli, and C. Ziras, "A review of data sources for electric vehicle integration studies," *Renewable and Sustainable Energy Reviews*, vol. 151, p. 111518, 2021, doi: 10.1016/j.rser.2021.111518.
[23] B. J. Mortimer, A. D. Bach, C. Hecht, D. U. Sauer, and R. W. de Doncker, "Public Charging Infrastructure in Germany — A Utilization and Profitability Analysis," *Journal of Modern Power Systems and Clean Energy*, pp. 1–10, 2021. [Online]. Available: https://ieeexplore.ieee.org/abstract/document/9582841
[24] C. Hecht, S. Das, C. Bussar, and D. U. Sauer, "Representative, empirical, real-world charging station usage characteristics and data in Germany," *eTransportation*, vol. 6, no. 4, p. 100079, 2020, doi: 10.1016/j.etran.2020.100079.
[25] ADAC, *Automarken & Modelle*. [Online]. Available: https://www.adac.de/rund-ums-fahrzeug/autokatalog/marken-modelle/ (accessed: Apr. 6 2020).
[26] Christopher Hecht, Jan Figgener, and Dirk Uwe Sauer, "ISEAview – Elektromobilität," 2021.
[27] Electric Vehicle Database, *Useable battery capacity of full electric vehicles*. [Online]. Available: https://ev-database.org/cheatsheet/useable-battery-capacity-electric-car (accessed: Mar. 14 2022).
[28] Statistisches Bundesamt Deutschland, *Experimentelle Daten - Mobilitätsindikatoren mit Mobilfunkdaten*. [Online]. Available: https://www.destatis.de/DE/Service/EXDAT/Datensaetze/mobilitaetsindikatoren-mobilfunkdaten.html (accessed: Jan. 27 2022).
[29] R. Follmer and D. Gruschwitz, "Mobilität in Deutschland – MiD Kurzreport. Ausgabe 4.0," Studie von infas, DLR, IVT und infas 360 im Auftrag des Bundesministers für Verkehr und digitale Infrastruktur, Bonn, 2019. Accessed: Jun. 22 2020. [Online]. Available: http://www.mobilitaet-in-deutschland.de/pdf/infas_Mobilitaet_in_Deutschland_2017_Kurzreport.pdf
[30] C. Hecht, J. Figgener, and D. U. Sauer, "Predicting Electric Vehicle Charging Station Availability Using Ensemble Machine Learning," *Energies*, vol. 14, no. 23, p. 7834, 2021, doi: 10.3390/en14237834.
[31] ADAC, *E-Ladesäulen: Parkbeschilderung of unklar*. [Online]. Available: https://www.adac.de/rund-ums-fahrzeug/elektromobilitaet/laden/parken-e-ladesaeulen/ (accessed: Mar. 14 2022).
[32] Federal Grid Agency and Federal Cartel Office, "Monitoringbericht 2021," Federal Grid Agency; Federal Cartel Office, Berlin, 2021. Accessed: Mar. 14 2022. [Online]. Available: https://www.bundesnetzagentur.de/SharedDocs/Mediathek/Monitoringberichte/Monitoringbericht_Energie2021.pdf?__blob=publicationFile&v=2
[33] Z. Crockett, *Why most gas stations don't make money from selling gas*. [Online]. Available: https://thehustle.co/why-most-gas-stations-dont-make-money-from-selling-gas/ (accessed: Mar. 21 2022).
[34] S. Horsley, *Gas Stations Profit from More Than Just Gas*. [Online]. Available: https://www.npr.org/templates/story/story.php?storyId=10733468 (accessed: Mar. 21 2022).
[35] EEA, *Corine Land Cover (CLC) 2018, Version 2020_20u1*. Brussels: European Environment Agency (EEA) under the framework of the Copernicus programme, 2018. Accessed: Jun. 24 2020. [Online]. Available: https://land.copernicus.eu/pan-european/corine-land-cover/clc2018
[36] C. Hecht, *BeNutz LaSA: Bessere Nutzung von Ladeinfrastruktur durch Smarte Anreizsysteme*. [Online]. Available: https://benutzlasa.de/ (accessed: Mar. 19 2021).
[37] Bundesamt für Kartographie und Geodäsie, *VG250*. © GeoBasis-DE / BKG 2020. [Online]. Available: https://gdz.bkg.bund.de/index.php/default/open-data/verwaltungsgebiete-1-250-000-ebenen-stand-01-01-vg250-ebenen-01-01.html (accessed: Apr. 2 2020).
[38] C. Olk, M. Trunschke, C. Bussar, and D. U. Sauer, Eds., *Empirical Study of Electric Vehicle Charging Infrastructure Usage in Ireland*. Xiamen, China: IEEE, 2019.

*A. Deciles of charge events by rated power*

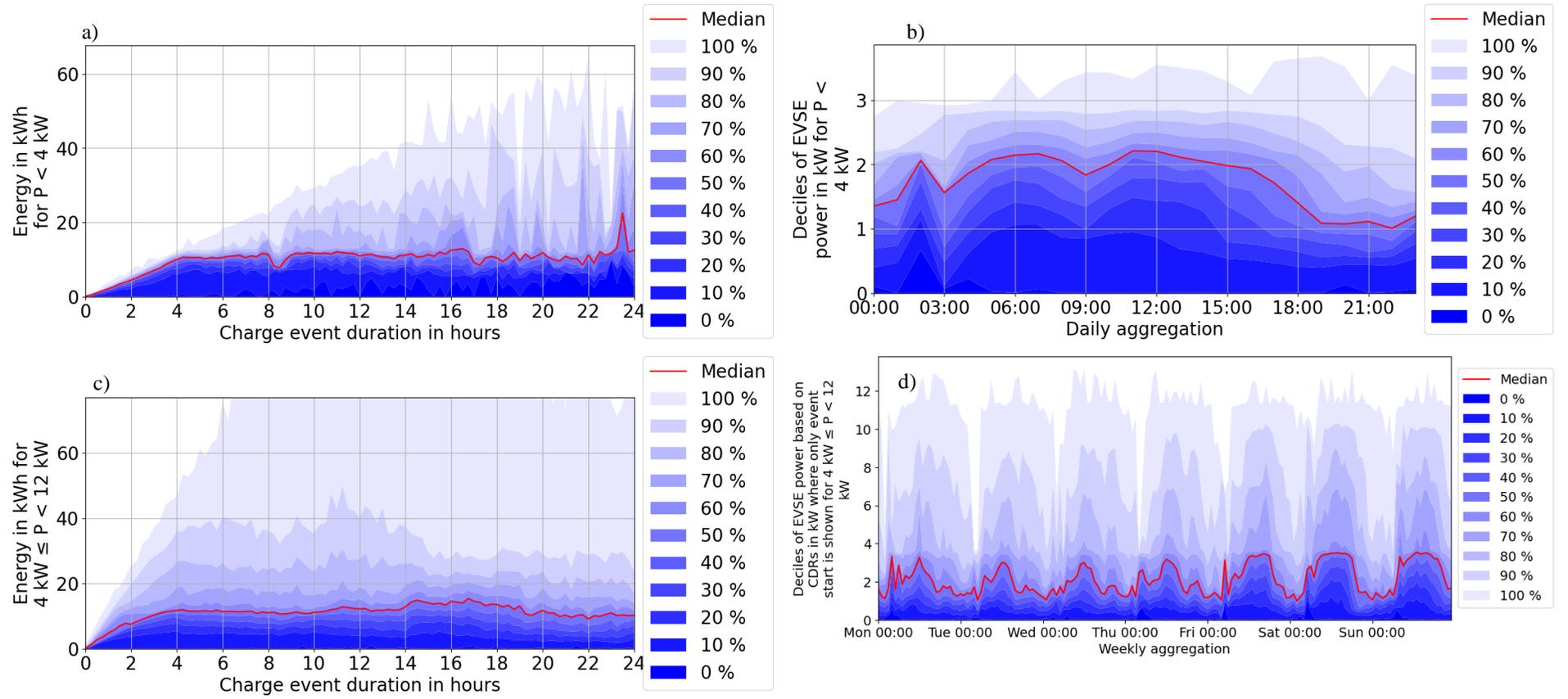

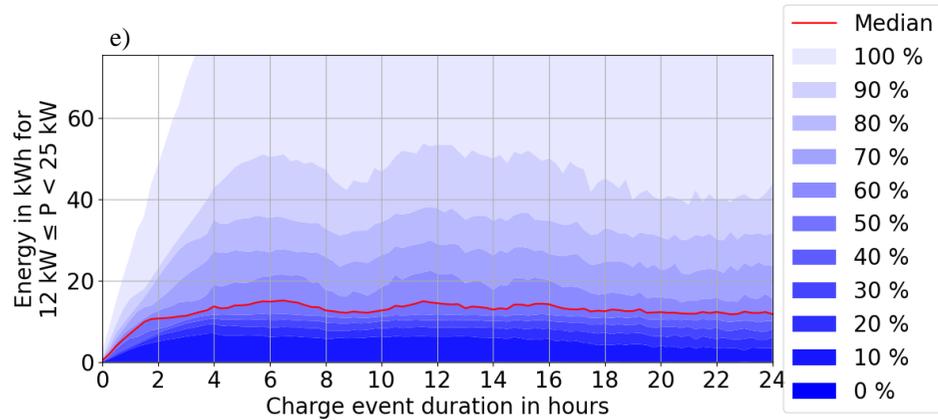
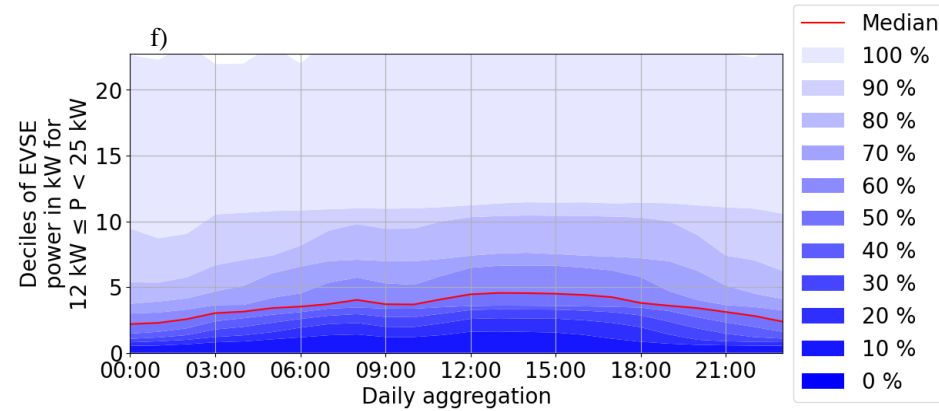
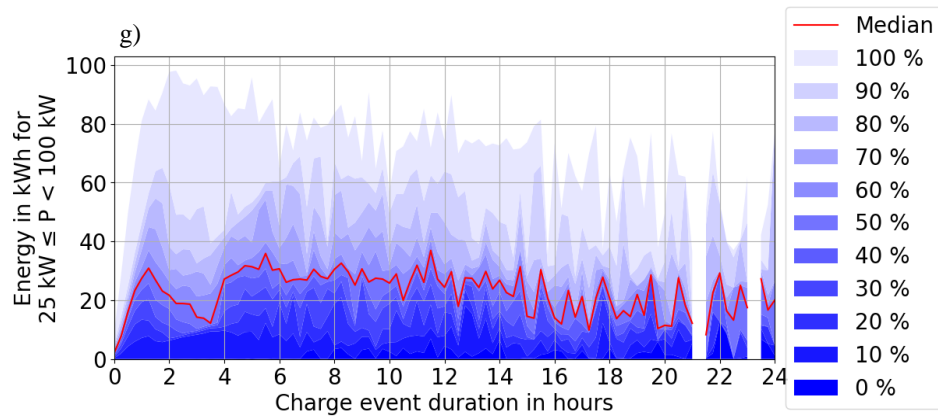
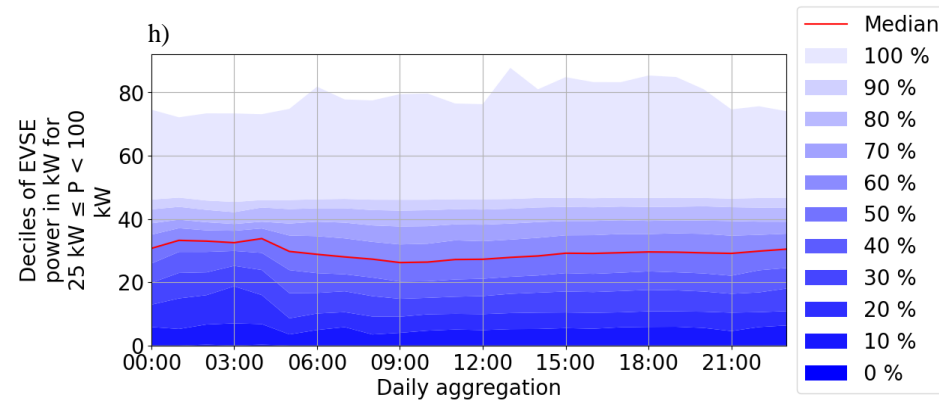
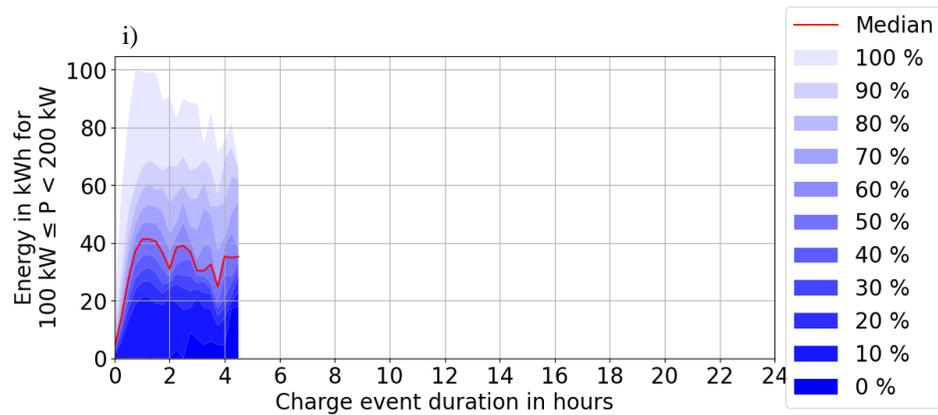
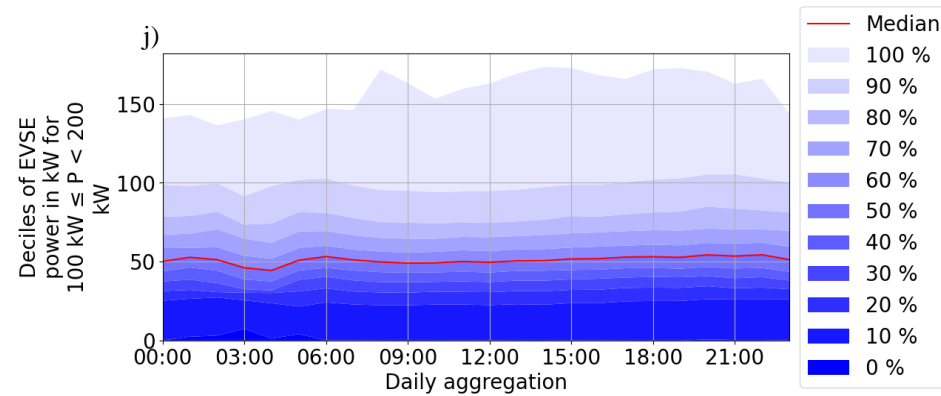

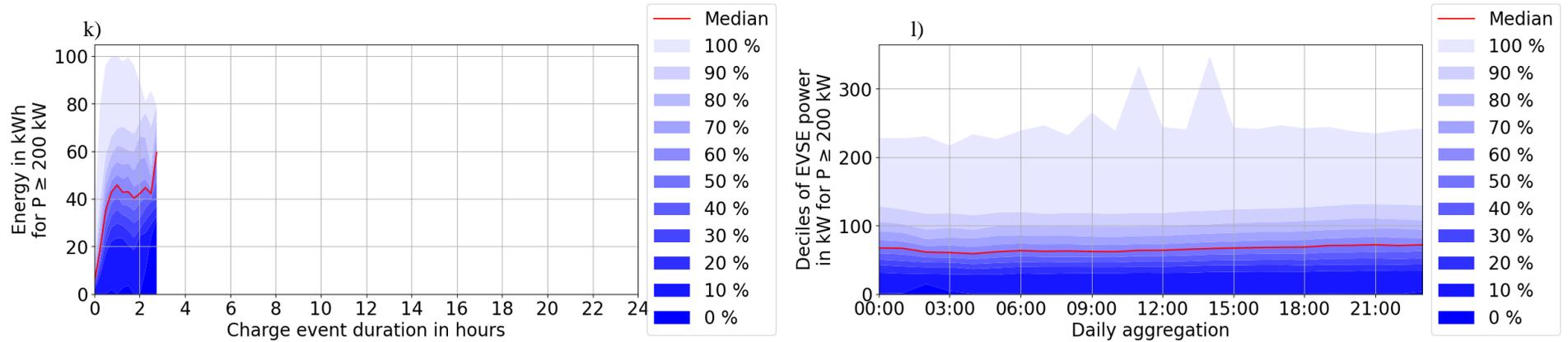

*Figure 10: Deciles of power flow for charge events recorded in the CDR data. The shade of blue indicates what percentage of events was below the fraction stated in the legend for each duration. The power ratings of the group of EVSEs for which events are displayed can be found on the y-axis of each plot with levels increasing in the standard order of this paper. For the daily aggregation shown on the right, the value plotted always corresponds to the power of events starting at the time and not over the entire event duration. Examples how to read: For charge events lasting around 15 minutes (with quarterhourly rounding windows) at EVSEs with a power rating above 200 kW, 80% of events were at a power level of less than ~73 kW when averaged over event duration (left plots). For charge events at EVSEs with a rated power between 12 and 25 kW and starting at midday, 50% experienced an average power of 4.45 kW over the entire event duration. Note that events lasting a time x do not influence the shown data for durations lower than x. Events lasting between 15 and 30 minutes consequently are not counted for the data point showing the power level of events lasting between 0 and 7.5 minutes. Also note that no values are displayed if less than 10 events with the given duration occurred explaining the large white spaces particularly for fast-chargers.*

### B. Meaning of different line styles in plots

Line plots in this paper are either solid, striped, dotted or blank. The definition of these styles is given in the table below.

*Table 4: Definition of line styles in all plots showing average values*

| Line style | Meaning |
| --- | --- |
| Wide and solid | The data points forming the line consist of at least 100 observations each. |
| Medium width and dashed | The data points forming the line consist of at least 20 observations each. |
| Thin and dotted | The data points forming the line consist of less than 20 observations each. |
| Blank | No line is shown if no corresponding data points are available |

Data points are generated in regular intervals across all plots these intervals generally are:

- every 15 minutes for plots showing a time window of 24 hours
- every 60 minutes for plots showing a weekly aggregation or no aggregation at all